\documentclass[aps,prb,twocolumn,floatfix,superscriptaddress,longbibliography,showpacs]{revtex4-1}
\usepackage[utf8]{inputenc}
\usepackage{graphicx}
\usepackage{dcolumn}
\usepackage{bm}
\usepackage{amsfonts}
\usepackage{amsmath}
\usepackage{amssymb}
\usepackage{braket}
\usepackage{color,soul}
\usepackage{wasysym}
\usepackage{mathrsfs}
\usepackage{times}
\usepackage[dvipsnames,svgnames,table]{xcolor}
\usepackage{ulem}
\usepackage{hyperref}
\usepackage{fancyhdr}
\usepackage[english]{babel}
\usepackage{hyperref}
\usepackage{float}
\usepackage{longtable}
\usepackage{multirow}

\hypersetup{
    unicode=true,          
    pdftoolbar=true,        
    pdfmenubar=true,        
    pdffitwindow=false,     
    pdfstartview={FitH},    
    pdftitle={2017-CONICYT},    
    pdfauthor={LEFFT},     	
    colorlinks=true,       	
    linkcolor=NavyBlue,  
    citecolor=Maroon,       
    filecolor=NavyBlue,      
    urlcolor=NavyBlue       
}

\newcommand{\acos}{\mathrm{acos}}
\newcommand{\ams}{\text{ }\textup{\AA}}

\begin{document}

\title{Robust edge states induced by electron-phonon interaction in graphene nanoribbons}

\author{Hern\'an L. Calvo}
\affiliation{Instituto de F\'isica Enrique Gaviola (CONICET) and FaMAF, Universidad Nacional de C\'ordoba, Argentina}
\affiliation{Departamento de F\'isica, Universidad Nacional de R\'io Cuarto, Ruta 36, Km 601, 5800 R\'io Cuarto, Argentina}

\author{Javier S. Luna}
\affiliation{Instituto de F\'isica Enrique Gaviola (CONICET) and FaMAF, Universidad Nacional de C\'ordoba, Argentina}

\author{Virginia Dal Lago}
\affiliation{Instituto de F\'isica Enrique Gaviola (CONICET) and FaMAF, Universidad Nacional de C\'ordoba, Argentina}

\author{Luis E. F. Foa Torres}
\affiliation{Departamento de F\'isica, Facultad de Ciencias F\'isicas y Matem\'aticas, Universidad de Chile, Santiago, Chile}

\begin{abstract}
The search of new means of generating and controlling topological states of matter is at the front of many joint efforts, including bandgap 
engineering by doping and light-induced topological states. Most of our understading, however, is based on a single particle picture. Topological 
states in systems including interaction effects, such as electron-electron and electron-phonon, remain less explored. By exploiting a 
non-perturbative and non-adiabatic picture, here we show how the interaction between electrons and a coherent phonon mode can lead to a bandgap 
hosting edge states of topological origin. Further numerical simulations witness the robustness of these states against different types of disorder. 
Our results contribute to the search of topological states, in this case in a minimal Fock space.
\end{abstract}

\maketitle

\section{Introduction}
\label{sec:intro}

The search of topological states of matter is now reshaping condensed matter physics.~\cite{hasan2010,ortmann2015} The pioneering works in the 
1980s~\cite{thouless1982,haldane1988} bloomed about 20 years later with the prediction~\cite{kane2005,bernevig2006} and discovery of topological 
insulators in two~\cite{koenig2007} and three dimensions.~\cite{hsieh2008} This field is today more active than ever, with new trends and discoveries 
expanding its frontiers. This includes, for example, the search for gapless but topological phases such as Weyl semimetals,~\cite{yan2017} topological 
states induced by time-dependent fields (the so-called Floquet topological insulators),~\cite{oka2009,lindner2011,sentef2015} time-dependent lattice 
distortions~\cite{iadecola2014} and, more recently, topological states in non-Hermitian systems.~\cite{shen2018,martinezAlvarez2018} Today, 
topological states have a main role in the global search for means of achieving on-demand properties.~\cite{basov2017}

However, most of the current understanding of topological states remains at the level of a single-particle. The effect of interactions, both on 
topological phases predicted on the basis of a single-particle picture or as a mean of inducing new ones, stands out as a major problem. Previous 
studies along this direction have shown that electron-phonon interaction can either suppress~\cite{reijnders2014} or even 
induce~\cite{garate2013,saha2014,kim2015,antonius2016} non-trivial topological phases as the temperature increases. But the interaction between 
electrons and coherent phonons can also induce dressed states (even in the low-temperature limit), thereby requiring a careful analysis of the 
excitation spectrum of the composed system (electron and phonons). This type of interaction typically requires going beyond the adiabatic limit and 
has been predicted to lead to a phonon-induced bandgap opening in carbon nanotubes.~\cite{foatorres2006,foatorres2008} Experiments have also evidenced 
a breakdown of the Born-Oppenheimer approximation in graphene with the same type of high-symmetry optical phonons.~\cite{pisana2007} The recent 
observation of chiral phonons in 2D materials~\cite{zhu2018} also adds much interest in the context of possible ARPES 
experiments.~\cite{huebener2018} Furthermore, other authors have put forward the possibility of using optical means to control the electron-phonon 
interaction.~\cite{dutreix2017}

Here we examine a model for electron-phonon interaction in a quasi-one dimensional system and show that it may lead to robust topological edge states 
in a sample that otherwise lacked them. Specifically, we consider a graphene nanoribbon in the presence of a strong electron-phonon interaction with a 
single high-symmetry optical phonon mode. By exploiting a Fock space picture incorporating non-perturbative and non-adiabatic 
effects,~\cite{anda1994,bonca1995} we find that at the center of the phonon induced bandgaps (located at half the phonon energy above the Dirac point) 
there are edge states of topological origin induced by the electron-phonon interaction. Furthermore, our numerical simulations show that these states 
remain robust to different types of disorder and ribbon geometries.

\section{Hamiltonian model and Fock space solution scheme}
\label{sec:method}

To investigate the effects of the electron-phonon (e-ph) interaction, we use the framework introduced in 
Refs.~[\onlinecite{anda1994}] and [\onlinecite{bonca1995}]. The purpose of this section is to write the system's Hamiltonian in a basis for the 
electron-phonon Fock space corresponding to a single electron plus the excitations of the phonon mode. Since the description is coherent and as such 
the quantum phases are fully preserved, the e-ph interaction does not produce any phase randomization. This approach has been used for a variety of 
problems including vibration assisted tunneling in STM experiments,~\cite{mingo2000} transport through molecules~\cite{ness1999,emberly2000} and 
resonant tunneling in double barrier heterostructures.~\cite{foaTorres2001}

Let us consider a tight-binding description of graphene nanoribbons (GNRs) through the Hamiltonian
\begin{equation}
\hat{H} = - \sum_{\braket{i,j}} \gamma_{ij} \hat{c}_i^\dag \hat{c}_j,
\label{eq:Ham_1}
\end{equation}
where the sum runs over nearest neighbor carbon atoms and $\gamma_{ij}$ represents the hopping amplitude connecting them. The fermion operator 
$\hat{c}_i^\dag$ ($\hat{c}_i$) creates (annihilates) an electron at site $i$ of the lattice. Carbon displacements $\delta \bm{r}_i$ from their 
respective equilibrium positions $\bm{r}_i^0$ are incorporated as a renormalization of the bare hopping amplitude 
$\gamma_0 = 2.7\text{ eV}$ through~\cite{pereira2009} $\gamma_{ij} = \gamma_0 \exp [-b (d_{ij}/a_0 - 1)]$, where 
$d_{ij}=|\bm{r}_i^0-\bm{r}_j^0+\delta\bm{r}_i-\delta\bm{r}_j|$ accounts for the distance between carbons $i$ and $j$, 
$a_0=|\bm{r}_i^0-\bm{r}_j^0|\simeq 1.42\ams$ is the equilibrium C-C distance, and $b \simeq 3.37$ is the rate of decay.

\begin{figure}[h]
\centering
\includegraphics[width=0.95\columnwidth]{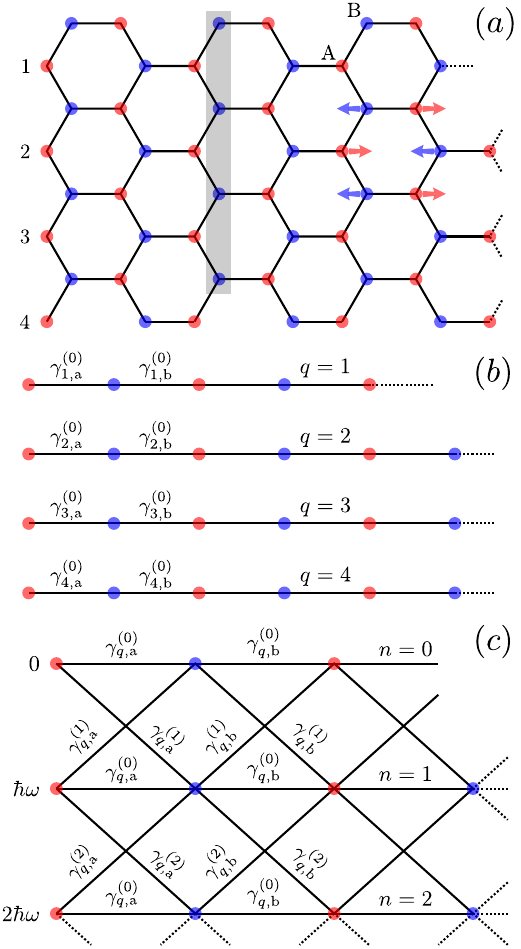}
\caption{Armchair GNR with phonon mode $A_1(L)$. (a) ac-GNR of $N_y=4$ carbon atoms wide ($L_y = 8.61\ams$). The shaded rectangle encloses a 
``transversal layer'', i.e. a single line of $N_y$ carbon atoms. The longitudinal optical phonon is depicted by red and blue arrows. (b) Eigenmode 
decomposition of the lattice. In the non-interacting case, each eigenmode consists of a dimer chain with intracell hoppings 
$\gamma_{q,\mathrm{a}}^{(0)}$ and intercell hoppings $\gamma_{q,\mathrm{b}}^{(0)}$. (c) Interacting case: Each dimer chain (eigenmode $q$), now 
splits into infinite replicas with different number of phonon excitations.}
\label{fig:1}
\end{figure}

As depicted in Fig.~\ref{fig:1}(a), we consider a single phonon mode characterized by a rigid displacement $\delta \bm{r} = a_0 Q \bm{u}$ between 
sublattices A and B, where $Q$ sets the strength of the displacement and $\bm{u} = (\cos\phi,\sin\phi)$ its direction.~\footnote{Throughout this work 
we will assume a longitudinal optical mode, such that $\bm{u}=(1,0)$.} The positions of the carbon atoms thus depend on which sublattice they 
belong, i.e. $\bm{r}_i = \bm{r}_i^0 \pm \delta\bm{r}/2$, for $i \in \{\text{A, B}\}$, respectively. Assuming small displacements, i.e. $Q \ll 1$, we 
linearize the hoppings as $\gamma_{ij} = \gamma_0 ( 1 - bQ \cos \alpha_{ij})$, where $\alpha_{ij}$ is the angle subtended by the C-C bond and the 
displacement direction. Now we impose quantization on the mechanical coordinate, such that the above hoppings introduce the electron-phonon 
interaction. The full Hamiltonian therefore reads
\begin{equation}
\hat{H} = \hat{H}_\mathrm{el} + \hat{H}_\mathrm{ph}-\sum_{\braket{i,j}} \gamma_x \cos \alpha_{ij} \hat{c}_i^\dag \hat{c}_j (\hat{a}^\dag+\hat{a}),
\label{eq:Ham_op}
\end{equation}
where $\hat{a}^\dag$ ($\hat{a}$) creates (annihilates) one phonon excitation of frequency $\omega$, and $\gamma_x$ sets the strength of the e-ph 
interaction. The pure electronic Hamiltonian is given by the C-C hoppings in equilibrium, and writes as in Eq.~(\ref{eq:Ham_1}) but with the 
replacement $\gamma_{ij} \rightarrow \gamma_0$. In Eq.~(\ref{eq:Ham_op}) we also included the phonon Hamiltonian, given by
\begin{equation}
\hat{H}_\mathrm{ph} = \hbar \omega \hat{a}^\dag \hat{a}.
\end{equation}
We work within a Fock space spanned by $\ket{i,n} = \ket{i} \otimes \ket{n}$ states, where $\ket{i}$ describes a single electron state (usually 
referred to the site basis), while $\ket{n}$ sets the number of phonon excitations in the lattice. In this basis, the Hamiltonian of 
Eq.~(\ref{eq:Ham_op}) can be represented as the following matrix:
\begin{equation}
\bm{H} = \bm{H}_0 \otimes \bm{1}_\mathrm{ph} + \bm{1}_\mathrm{el} \otimes \hbar \boldsymbol{\Omega} + \bm{H}_1 \otimes \bm{X}.
\label{eq:Hfock}
\end{equation}
Here, $\bm{H}_0$ and $\bm{H}_1$ are electronic matrices representing the $\gamma_0$ and $\gamma_x$ hoppings in the hexagonal graphene lattice, 
$\bm{1}_\mathrm{ph}$ and $\bm{1}_\mathrm{el}$ are the identity matrices in phonon and electron subspaces, respectively, while the remaining phonon 
matrices are defined as:
\begin{equation}
\boldsymbol{\Omega} = 
\left(\begin{matrix}
0 & 0      & 0 	     &        \\
0 & \omega & 0 	     &        \\
0 & 0      & 2\omega &        \\
  &        &	     & \ddots
\end{matrix}\right), \;
\bm{X} = 
\left(\begin{matrix}
0        & \sqrt{1} & 0	       & \\
\sqrt{1} & 0        & \sqrt{2} & \\
0        & \sqrt{2} & 0        & \\
&        &          & \ddots
\end{matrix}\right).
\end{equation}

Expressed in the Fock space basis, the Hamiltonian in Eq.~(\ref{eq:Hfock}) can be visualized as the original one without interactions together with the 
replicas corresponding to different phonon excitations and the interactions between them. The presence of these excitations is accounted for by the 
additional energies $n\hbar\omega$. In this representation, the e-ph coupling enters through the last term, which enables the absorption and emission 
of a single phonon each time the electron `hops' between two carbon atoms.

\section{Electron-phonon induced edge states}
\label{sec:eph-on}

As introduced in the previous section, here we consider a fully quantized vibrational mode of frequency $\omega$, and describe the vibrating 
nanoribbons through a Fock space spanned by the states $\ket{i,n}$, which accounts for both the electronic and vibrational degrees of freedom. 

This might remind the reader of a similar picture used for time-periodic Hamiltonians: Floquet theory.~\cite{kohler2005} Indeed, there are several 
parallels stemming from a seeming isomorphism between the Floquet space and the Fock space,~\cite{shirley1965} but a few crucial differences must 
be noticed: (\textit{i}) Unlike for the case of time-dependent potentials, for the case of phonons, temperature plays a natural role in defining 
the phonon population. (\textit{ii}) In Floquet theory, the replica index is unbound while in the case of phonons described here it is bounded 
from below ($n \geq 0$). (\textit{iii}) The matrix elements for phonon emission and absorption change with the phonon population. Both descriptions 
(i.e. Fock space and Floquet space) match only when the system is in a highly excited state, a fact that is far from correct for optical phonons with 
typical energies exceeding $k_B T$ at room temperature. Aside from these differences, inspired from what we learned from Floquet topological 
states,~\cite{oka2009,lindner2011,perez-piskunow2014} one might then search for similar physics induced by the electron-phonon interaction. 

Another issue one might notice is the role of time-reversal symmetry (TRS): While a gap in irradiated graphene requires circularly 
polarized light so as to break TRS,~\cite{oka2009,calvo2011}~\footnote{We note the laser-induced bandgap opening has been experimentally 
observed through ARPES at the surface of a three-dimensional topological insulator~\cite{wang2013}.} the phonons considered here do not break 
such a symmetry (though they do open a bandgap~\cite{foatorres2006,foatorres2007,foatorres2008} at $\hbar \omega/2$ in the bulk material). 
However, one needs also to point out that in this work we are restricting ourselves to a ribbon geometry (i.e. a quasi one-dimensional system) rather 
than a two-dimensional system. Interestingly, chiral phonons,~\cite{zhang2015,zhu2018} lying at the corners of the Brillouin zone, could be used in 
two-dimensional hexagonal lattices to break TRS (at least locally in the valleys). Carrying on with the Fock-Floquet analogy, one could expect in the 
latter case similar physics as that of irradiated graphene with circularly polarized light.

\subsection{Vibration induced bandgaps}

The proposed vibration of the lattice consists in a single mode characterized by a rigid displacement between the two 
sublattices. We will work in the case where the displacement direction coincides with the longitudinal direction of the ribbon, i.e. $\phi=0$, 
motivated by the strong e-ph coupling observed in the optical mode $A_1(L)$ in CNTs leading to a Peierls-like 
mechanism~\cite{dubay2002,dubay2003,connetable2005} and Kohn anomalies;~\cite{farhat2007} and also in graphene samples.~\cite{pisana2007}

To begin with, we consider a graphene nanoribbon with armchair edge geometry (ac-GNR). The reason of this particular choice rests in the possibility 
it offers to decompose the system into a series of decoupled eigenmodes. To do so, we start from the non-interacting case ($\gamma_x = 0$) and we 
use the basis transformation proposed in Ref.~[\onlinecite{rocha2010}]. This transformation takes the ac-GNR into a series of $N_y$ eigenmodes, 
each one consisting on a dimer chain with alternating hoppings $\gamma_{q,\mathrm{a}}^{(0)} = 2 \gamma_0 \cos [\pi q/(2N_y+1)]$ and 
$\gamma_{q,\mathrm{b}}^{(0)} = \gamma_0$, with $q = 1,\dots,N_y$ the mode number [see Fig.~\ref{fig:1}(b)]. In this sense, one can identify each 
eigenmode $q$ as an independent Su-Schrieffer-Heeger (SSH) model,~\cite{su1979} for which the topological properties are 
well-known.~\cite{fefferman2014,li2014,asboth2016} 
Furthermore, the SSH model for time-dependent hoppings was also investigated in the context of Floquet topological states, both 
theoretically~\cite{gomez-leon2013,asboth2014,dallago2015,zhu2018a} and experimentally.~\cite{cheng2018}

When including the e-ph interaction, we can extend this mode decomposition in the Fock space, such that for each eigenmode we obtain a series
of dimer chains (or replicas of the non-interacting case), each one belonging to a different number $n$ of phonon excitations, see 
Fig.~\ref{fig:1}(c). This is easy to see regarding the structure of the Fock Hamiltonian in Eq.~(\ref{eq:Hfock}), where $\bm{H}_0$ and $\bm{H}_1$ 
commute with each other for this particular ribbon geometry and phonon mode. So, for a given mode $q$, the intrachain hoppings alternate between 
$\gamma_{q,\mathrm{a}}^{(0)}$ and $\gamma_{q,\mathrm{b}}^{(0)}$, as in the non-interacting case. The phonon energy $n \hbar \omega$ in the $n$-replica 
enters as a site energy along the whole chain, according to the second term in the r.h.s. of Eq.~(\ref{eq:Hfock}); and the interchain hoppings 
connecting the $n-1$ and $n$ replicas are given by $\gamma_{q,\mathrm{a}}^{(n)} = \sqrt{n} \gamma_x \cos[\pi q/(2N_y+1)]$ and 
$\gamma_{q,\mathrm{b}}^{(n)}=- \sqrt{n} \gamma_x$.

The degree of complexity imposed by the interaction clearly difficults the possibility of having analytic solutions for the system. However, the 
assumed weak coupling between the replicas ($\gamma_x \ll \gamma_0$) allows us to estimate the effects of the vibration on the electronic band 
structure by using perturbation theory around half the phonon energy. Other quantities like the local density of states (LDoS), eigenenergy spectrum 
and wavefunction amplitudes will be addressed numerically through standard techniques. 

To begin with, we neglect the coupling $\gamma_x$ between the different phonon replicas, such that the $(q,n)$-dispersion relation at zeroth order can 
be written as:
\begin{equation}
\varepsilon_{q,\pm}^{(n)}(k) =  n \hbar \omega \pm \sqrt{{\gamma_{q,\mathrm{a}}^{(0)}}^2 + {\gamma_{q,\mathrm{b}}^{(0)}}^2 + 
2 \gamma_{q,\mathrm{a}}^{(0)} \gamma_{q,\mathrm{b}}^{(0)} \cos ka}.
\label{eq:mode-dispersion-vib}
\end{equation}
This equation represents the band dispersion one would obtain for a dimer chain with unit cell length $a=3a_0/2$, but shifted in an integer number of 
$\hbar\omega$, which accounts for the phonons energy. Among the infinite number of replicas, we will take as reference the lowest energy bands 
belonging to the zero-phonon subspace. This is justified for optical phonons with energies largely exceeding the thermal energy at room temperature 
(the stretching mode in graphene has a phonon energy of about $200$ meV, this is about $8$ times the thermal energy at $300$K). Hence, the phonon 
population is zero for practical purposes. According to Eq.~(\ref{eq:mode-dispersion-vib}), the number of replicas crossing each other will depend on 
the relation between the phonon energy $\hbar\omega$ and the bandwidth associated with each replica, given by $6\gamma_0$ for $N_y > 1$. 

For $\hbar\omega \gg \gamma_0$, the distance between bands belonging to different replicas is much larger than their widths and, therefore, they do 
not cross each other. In this limit, the perturbation on the zero-phonon replica due to replicas with a higher number of phonons becomes negligible: 
The electron lying in a lattice without phonons can never reach enough energy as to spontaneously emit a phonon. 

The symmetry between the conduction ($+$) and valence $(-)$ bands with respect to $n\hbar\omega$ for $n=0$ and $n=1$ ensures that the band crossing 
takes place at half the phonon energy, $\varepsilon=\hbar\omega/2$. This fact, together with the maximum allowed energy for the zero-phonon band, 
imposes the condition $\hbar\omega<6\gamma_0$ for the first band crossing between two different phonon replicas. In general, for smaller values of 
$\omega$, the zero-phonon replica will cross with the $n$ phonon replica once the condition $n\hbar\omega<6\gamma_0$ is fulfilled.

\begin{figure}[t]
\includegraphics[width=\columnwidth]{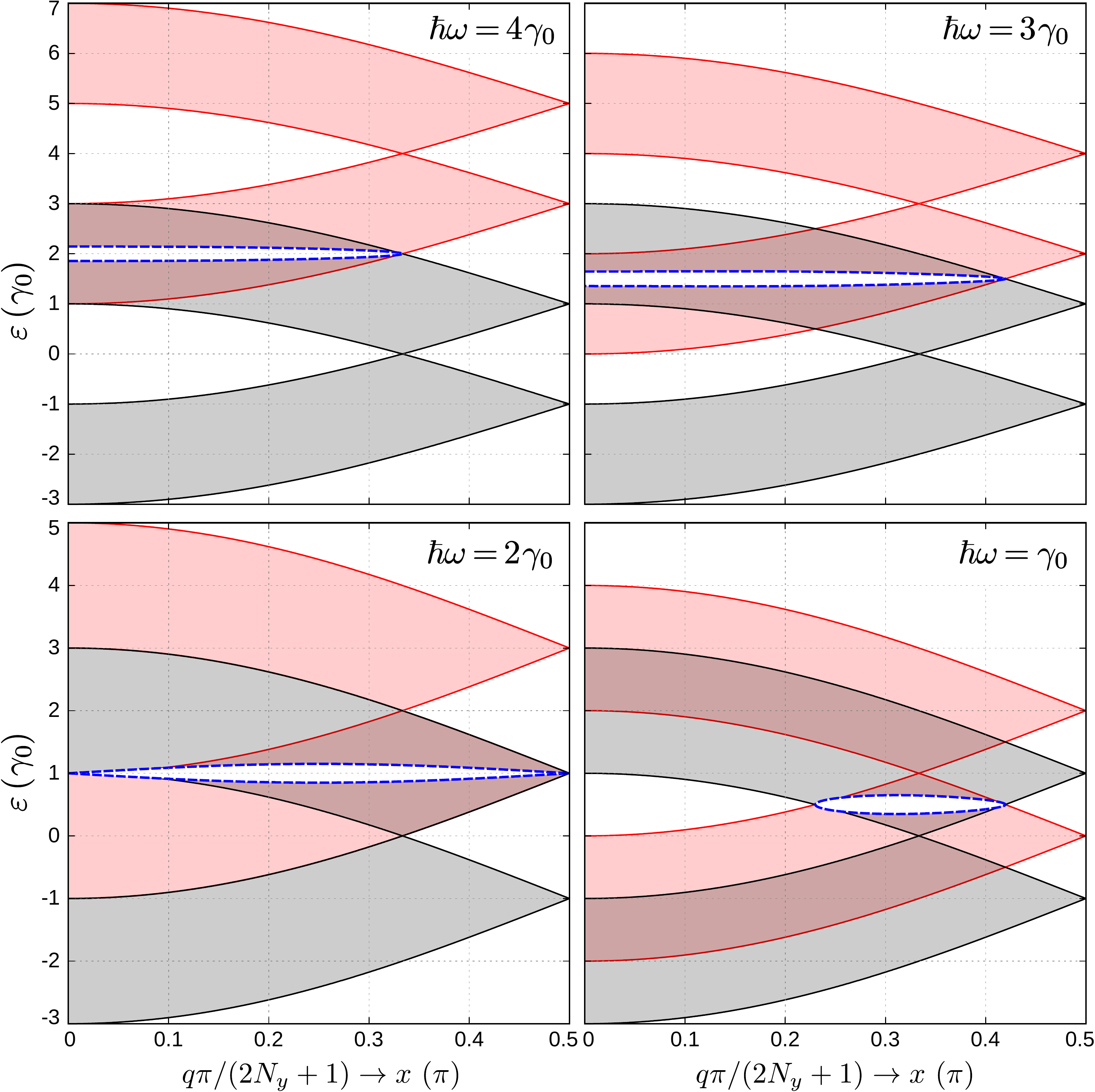}
\caption{Bandgap estimation according to Eq.~(\ref{eq:gaps}) as a function of the eigenmode number in the limit $N_y \rightarrow \infty$, for 
several values of the phonon frequency, and $\gamma_x=0.1\gamma_0$. The $n=0$ and $n=1$ phonon bands are shown as shaded regions in black and 
red, respectively. The dashed blue lines delimit the gapped regions.}
\label{fig:2}
\end{figure}

The eigenmode decomposition allows us to simplify the analysis around the band crossing processes. As in this particular geometry they are 
decoupled, the only relevant crossings are those with the same value of $q$. We can think, then, in the crossing between $n=0$ and $n=1$ bands 
belonging to the eigenmode $q$. To have such a crossing, it is necessary that the $n=0$ conduction band and the $n=1$ valence 
band overlap, which imposes the following range:
\begin{equation} 
|\gamma_0-\gamma_{q,\mathrm{a}}^{(0)}|<\frac{\hbar\omega}{2}<\gamma_0+\gamma_{q,\mathrm{a}}^{(0)}.
\label{eq:ranw}
\end{equation}
If such inequality can not be met, then the bands do not cross each other, and the zero-phonon replica gets virtually unperturbed. Conversely, if such 
inequality is fulfilled, the bands will cross at the $k$-points determined by the condition $\varepsilon_{q,+}^{(0)}(k) = 
\varepsilon_{q,-}^{(1)}(k)$. This equation has two solutions, $\pm k_q^*$, due to the symmetric dispersion of the bands in 
Eq.~(\ref{eq:mode-dispersion-vib}) around $k=0$. When we include the e-ph interaction through $\gamma_x$, the crossing between the bands gets 
avoided, yielding a gap induced by the vibration. For these $k_q^*$ the group velocity goes to zero, meaning that a new backscattering process was 
introduced. From the point of view of the $n=0$ replica, an electron traveling in a static ribbon~\footnote{We here neglect the zero point motion of 
the lattice} with energy $\varepsilon_{q,+}^{(0)}(k_q^*)$ may suffer a reflection, together with the emission of a phonon of energy $\hbar\omega$. 
Alternatively, from the point of view of the $n=1$ replica, an electron traveling in a vibrating ribbon, such that the composite e-ph system has 
energy $\varepsilon_{q,-}^{(1)}(k_q^*)$, may also be reflected, after absorption of the phonon excitation present in the lattice. 

Supposing the limit case $N_y \rightarrow \infty$, we can take the cosine argument in $\gamma_{q,\mathrm{a}}^{(0)}$ as a continuous variable 
$x=q\pi/(2N_y+1)$ in the range $0 < x < \pi/2$. Under the perturbative regime, we estimate the bandgap size from a reduced Hamiltonian which only 
includes those bands which are expected to cross (see App.~\ref{app:1}). The size of the gap, as a function of $x$, writes:
\begin{equation}
\Delta(x)=\frac{3\gamma_x}{\eta}\sqrt{\left(\eta_{+}^2-\cos^2 x\right)\left(\cos^2 x-\eta_{-}^2\right)}, 
\label{eq:gaps} 
\end{equation}
where $\eta=\hbar\omega/4\gamma_0$ and $\eta_{\pm} = \eta \pm 1/2$. In Fig.~\ref{fig:2} we show this vibration induced bandgap for several values of 
the phonon frequency in the limit $N_y \rightarrow \infty$. To some extent, one can regard this limit as taking the nanoribbon into a two-dimensional 
graphene layer. Here, the set of eigenmodes becomes dense, meaning that there is always a specific $x^\ast$-value in which the vibration 
induced gap necessarily closes. This is a consequence of the preserved TRS by the phonon mode. Going back to the finite $N_y$ system, the set of 
$x$-values is no longer dense, and the overall gap will be given by the $q$-mode closest to $x^\ast$. This somewhat relaxes the need to break TRS as 
to open a gap in quasi one-dimensional systems.

We notice that Eq.~(\ref{eq:gaps}) has physical meaning as long as the argument of the square root is positive. This sets the condition for those 
eigenmodes in which there is a band crossing, and it results to be $0<x<\acos(\eta_{-})$ for $\eta>1/2$ and 
$\acos(\eta_{+})<x<\pi-\acos(\eta_{-})$ for $\eta<1/2$, in agreement with Eq.~(\ref{eq:ranw}). On the other hand, the bandgap size depends linearly on 
$\gamma_x$, and there is also some dependence with the phonon frequency through the $\eta$ parameter. Interestingly, while for semiconducting GNRs 
the bandgap around the central region $\varepsilon=0$ can only be closed by increasing $N_y$, the vibration induced bandgap could be controlled, to 
some extent, through the modulation of the e-ph coupling.~\cite{dutreix2017} We can, in turn, determine the maximum value of the gap as a function of 
$\hbar\omega$. This yields two regimes: (\textit{i}) for $\eta<\sqrt{3}/2$, a frequency independent regime with maximum gap $\Delta_{\max}=3\gamma_x$, 
and (\textit{ii}) for $\eta>\sqrt{3}/2$, a frequency dependent regime where the maximum gap decreases and it closes in the limit 
$\hbar\omega=6\gamma_0$.

\subsection{Spectral properties and characterization of the edge states}

Throughout the following analysis we will consider a high-frequency regime by choosing $\hbar\omega=5\gamma_0$.~\footnote{For the chosen value 
$\hbar\omega=5\gamma_0$, the band crossings occur for those eigenmodes fulfilling $q < q_{\max} = (2N_y+1) \, \acos(3/4)/\pi$. } For this 
value, the conduction band belonging to the zero-phonon replica only cross with the valence band of the one-phonon replica, since the condition 
$n\hbar\omega < 6\gamma_0$ can only be fullfilled by $n=1$. Although the chosen $\hbar\omega$ largely exceeds the typical phonon energy of the optical 
mode, this high-frequency regime, together with the weak e-ph coupling assumption ($\gamma_x \ll \gamma_0$), allows us to truncate the infinite Fock 
space in the first two phonon replicas, thereby simplifying the discussion of the vibration effects on the electronic properties of the ribbon. In 
particular, this ensures both the valence band ($\varepsilon<0$) and the semiconducting gap ($\varepsilon = 0$) regions being unaffected by the 
vibration, at least in lowest order in the e-ph coupling. This allows us to establish a clear distinction between the well-known ``native'' 
edge-states~\cite{delplace2011} at $\varepsilon = 0$, and the expected e-ph induced edge-states, located at $\varepsilon = \hbar\omega/2$. In any 
case, the same analysis can be carried out for $\hbar\omega \simeq 200$ meV, but keeping in mind that a competition between the native topology and 
the e-ph induced edge states may occur near the charge neutrality point.~\cite{dallago2015}

\begin{figure}
\includegraphics[width=\columnwidth]{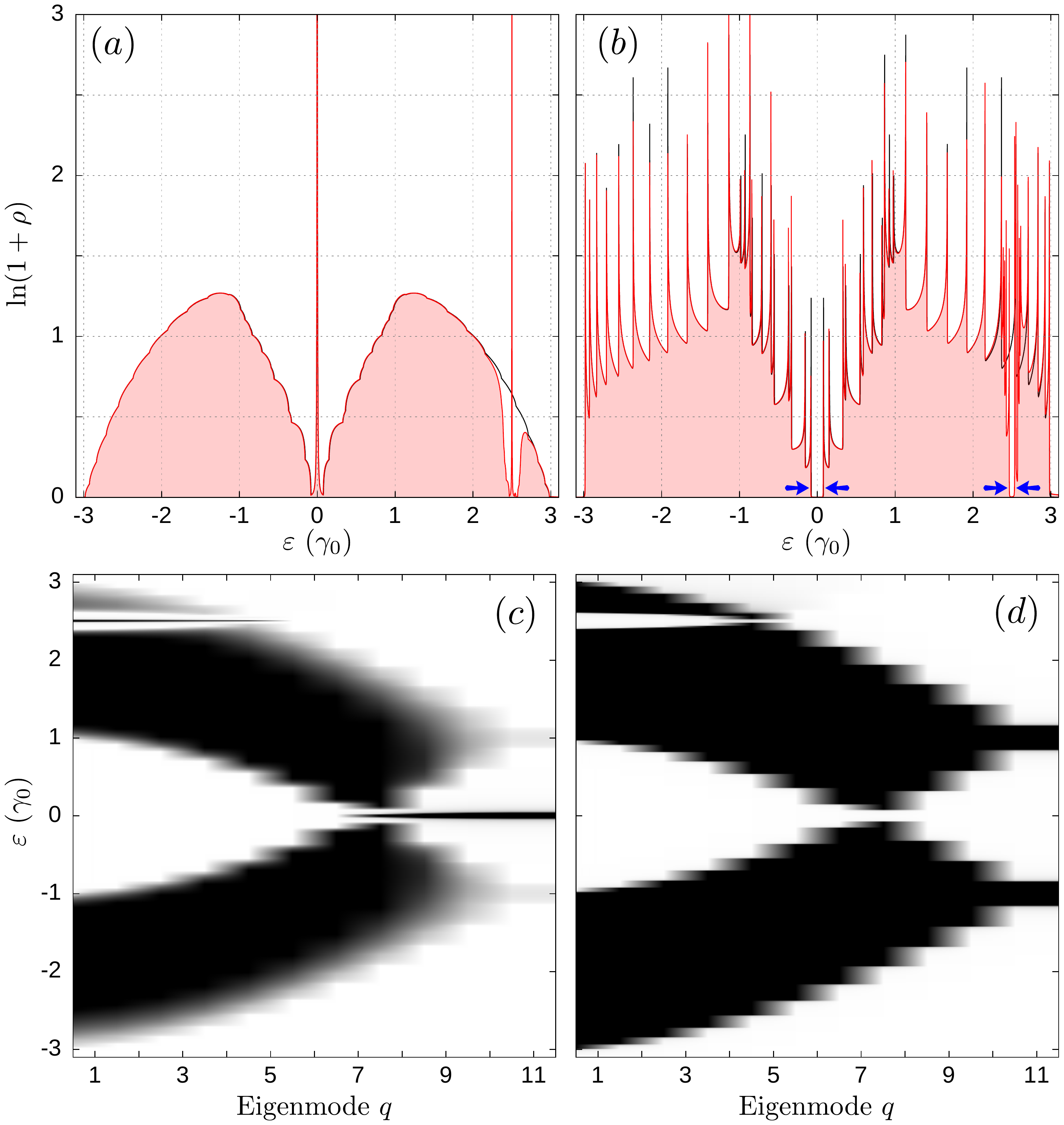}
\caption{LDoS plots for an ac-GNR of $N_y=11$ carbon atoms wide ($L_y = 25.82 \ams$) and coupled to a longitudinal optical mode. (a) LDoS in log 
scale weighted over the zero-phonon replica (solid red, shaded area), evaluated at the border of a semi-infinte ac-GNR. (b) Zero-phonon LDoS in bulk 
situation, i.e. evaluated at the center of an infinite ac-GNR. In (a) and (b) the black lines show the non-interacting case ($\gamma_x = 0$). Panels 
(c) and (d) show the same densities (in grayscale) as in panels (a) and (b), respectively, but resolved in eigenmodes. The vibration parameters are 
$\hbar\omega=5\gamma_0$ and $\gamma_x=0.1\gamma_0$.}
\label{fig:3}
\end{figure}

In Fig.~\ref{fig:3}(a) we show the LDoS weighted over the zero-phonon replica (red shaded area), evaluated at the lateral border of a semi-infinite 
ac-GNR of $N_y = 11$ carbon atoms wide ($L_y = 25.82 \ams$). Except for some region around $\varepsilon=\hbar\omega/2$, the shape of the LDoS is 
qualitatively the same as that of the non-interacting case (solid black), with a central peak related with the native edge states. This is expected as 
the e-ph interaction only becomes effective in the band crossing region at $\hbar\omega/2$. Here, we observe a similar behavior as in the 
$\varepsilon=0$ region, i.e. a depletion in the LDoS with a pronounced peak rising in its center. 

To infer whether the peak at $\hbar\omega/2$ survives far away from the border of the ribbon, we show in Fig.~\ref{fig:3}(b) the 
zero-phonon LDoS evaluated at the center of an infinite ac-GNR. The pronounced peaks at $\varepsilon = 0$ and $\hbar\omega/2$ are no longer 
visible, and instead we can observe bandgap openings around these energies (blue arrows). For $\hbar\omega/2$ this is the e-ph interaction induced 
bandgap, which was also predicted in vibrating CNTs.~\cite{foatorres2006,foatorres2007,foatorres2008} As it is expressed in Eq.~(\ref{eq:gaps}), the 
size of the vibration induced gap depends on the eigenmode we are looking at, which is observed in the LDoS maps of Figs.~\ref{fig:3}(c) 
and (d). For the chosen values $\hbar\omega=5\gamma_0$ and $N_y = 11$, the band crossings occur for those eigenmodes fulfilling 
$q < q_{\max} \simeq 5.29$. The minimum gap occurs for the mode with $q$ closest to $q_{\max}$, which in this case is $q = 5$. Here, the effective 
interchain coupling between the phonon replicas [c.f. Eq.~(\ref{eq:Ham_eff})] is the smallest, and it increases for smaller $q$ modes.

\begin{figure*}
\includegraphics[width=\textwidth]{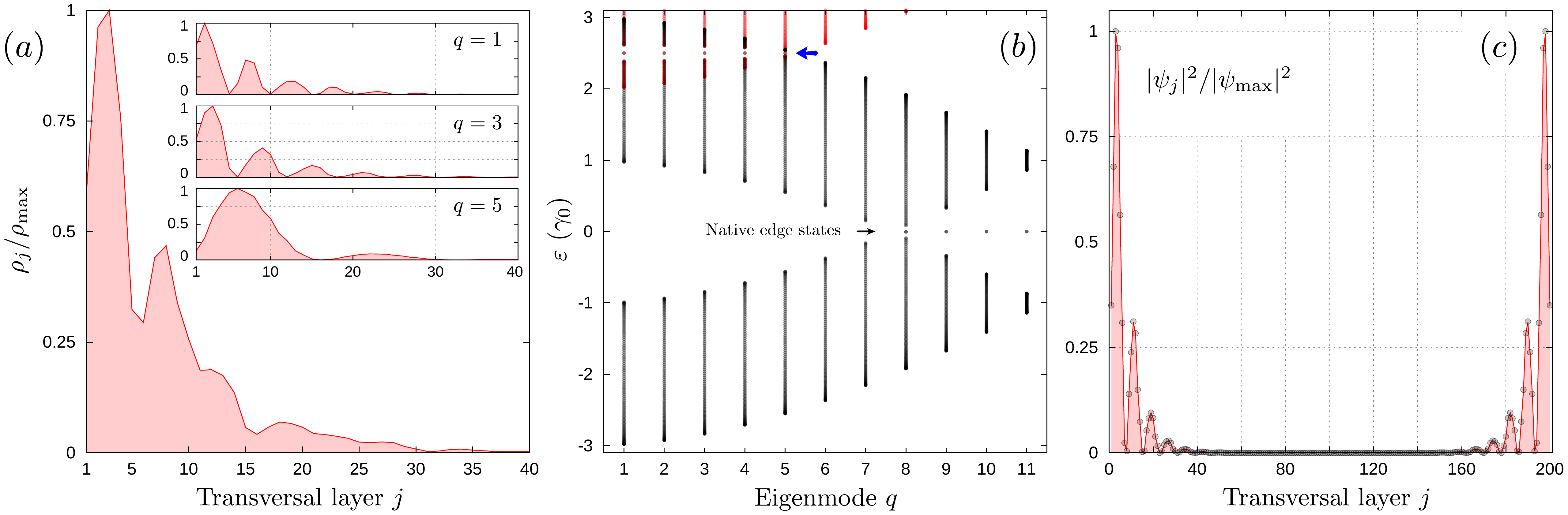}
\caption{(a) LDoS evaluated at $\varepsilon=\hbar\omega/2$ as a function of the transversal layer $j$ for a semi-infinite ac-GNR with $N_y = 11$ 
carbon atoms wide and normalized to its maximum value at $j=3$. The used phonon energy is $\hbar\omega=5\gamma_0$ and the e-ph interaction is 
$\gamma_x = 0.1$. Inset: Same LDoS as in (a), for eigenmodes $q=1$, $3$, and $5$ and normalized to their respective maximum 
values. (b) Eigenenergy spectrum for a finite ac-GNR of dimensions $N_x=200$ ($L_x=211.58 \ams$) and $N_y=11$ ($L_y = 25.82 \ams$). The other 
parameters coincide with those of panel (a). We use a color scale to indicate the eigenstates weight with respect to the zero-phonon replica: Full 
weight ($p_{q,k}=1$) is in black, while zero weight ($p_{q,k}=0$) is in red. (c) Probability density $|\psi_j|^2$ (normalized to its maximum value 
and weighted over the $n=0$ replica) as a function of the transversal layer $j$ for one of the two eigenenergies at $\varepsilon = \hbar\omega/2$ and 
eigenmode $q = 4$ in panel (b).}
\label{fig:4}
\end{figure*}

The absence of peaks in the bulk LDoS makes us suspect that, as in the case of the native edge states at $\varepsilon = 0$, the peak at 
$\hbar\omega/2$ is also related with states localized at the ribbon's border. Let us see, now, the behavior of the vibration induced peak as we 
move inside the ribbon. In Fig.~\ref{fig:4}(a) the zero-phonon LDoS evaluated at half the phonon energy is shown as a function of the transversal 
layer number $j$ [vertical lines of carbon atoms, see Fig.~\ref{fig:1}(a)]. The LDoS decays exponentially, such that for $j \sim 40$ it becomes 
negligible. The way in which the peak decays is quite irregular, due to the superposition of the contributing tranversal modes (see inset 
plots in the figure).

To characterize the eigenenergy spectrum and the localized state wavefunctions we now consider a finite ac-GNR of $N_x = 200$ carbon atoms long 
($L_x = 211.58\ams$) and $N_y = 11$ wide ($L_y=25.82\ams$). In Fig.~\ref{fig:4}(b) we show the eigenenergy spectrum resolved in eigenmodes. To 
visualize such a spectrum as a perturbation of that in the non-interacting case, we weight each eigenstate over the $n=0$ replica, i.e. the 
zero-phonon subspace. Given a $k$-eigenstate belonging to the eigenmode $q$, we can decompose it as the following superposition among the $n$ 
replicas according to:
\begin{equation}
\ket{\psi_{q,k}} = \sum_n\sum_j \braket{\varphi_{q,j}^{(n)}|\psi_{q,k}} \ket{\varphi_{q,j}^{(n)}} = \sum_n \ket{\psi_{q,k}^{(n)}},
\end{equation}
and take its projection over the $n = 0$ subspace, i.e.
\begin{equation}
p_{q,k} = |\braket{\psi_{q,k}^{(0)}|\psi_{q,k}}|^2.
\end{equation}
Although the number of replicas is infinite, the fact that we work in a perturbative regime allows us to truncate the full Fock space in a few 
replicas. In our case where $\hbar\omega = 5\gamma_0$ and $\gamma_x \ll \gamma_0$, the subspace associated with those replicas with $n>1$ has 
negligible impact on the $n=0$ replica. In Fig.~\ref{fig:4}(b), the red dots reveal how the valence band belonging to the $n=1$ replica mixes with the 
conduction band of the $n=0$ replica (black dots) for $\varepsilon \sim \hbar\omega/2$. Around this value, we can observe bandgap openings for those 
modes with $q < q_\mathrm{max}$, together with the presence of two degenerate midgap states per eigenmode (blue arrow). 

In Fig.~\ref{fig:4}(c) we show one of the two midgap states for $q = 4$. As we move towards the center of the ribbon, the probability density 
oscillate according with the weights this particular $q$-mode has on the sites composing the transversal layer. As it happens with the LDoS peak, the 
wavefunction also shows an exponential decay along the longitudinal direction, with a typical ‘inverse gap’ localization length. However, as the 
ribbon in this example has a finite length, the wavefunction has weight in the two borders. We can think of this state as bonding or antibonding 
combination of two states, $\ket{\psi_L}$ and $\ket{\psi_R}$, localized at the left and the right border of the ribbon, respectively. For a ribbon 
with a large number of transversal layers (as it happens here), the overlap $\braket{\psi_L|\psi_R}$ can be neglected and one can consider 
$\ket{\psi_L}$ and $\ket{\psi_R}$ as linear combinations of the two degenerate midgap states.

\subsection{Topological origin of the edge states}
\label{sec:topo-2}

As discussed before, when decreasing the phonon energy from $\hbar\omega = 6 \gamma_0$ on we will observe band crossings between different phonon 
replicas which, in turn, generate bandgap openings as new backscattering processes are introduced. This can be regarded as a band inversion process: 
In the crossing region, the valence band from the $n=1$ replica happens to have less energy than that of the conduction band from the $n=0$ replica. 
This band inversion is characteristic in topological phase transitions, together with the formation of localized midgap states. This strongly 
motivates the calculation of the Zak phase to infer about the topological nature of the vibration induced localized states. Although there is an 
infinite number of bands due to the structure of the Fock space, we can again truncate this by taking only those bands belonging to the $n=0$ and 
$n=1$ replicas. The corresponding topological invariant can be calculated as the integral of the Berry connection~\cite{zak1989}
\begin{equation}
\mathcal{Z}_\alpha = i \oint dk \braket{u_{k,\alpha}|\partial_k u_{k,\alpha}},
\end{equation}
with $\ket{u_{k,\alpha}}$ the Bloch states belonging to the $\alpha$-band and the integral taken over the first Brillouin zone. The bulk-boundary 
correspondence then allows us to characterize the existence of topological states with the Zak phase. Summing up 
$\mathcal{Z}_\alpha$ for all the bands with energy below a given gap yields the cumulative phase (modulo $2\pi$), which indicates the existence 
(with cumulative phase $\pi$) or absence (zero cumulative phase) of topological midgap states.

Although the Zak phase can be obtained analytically in the SSH model~\cite{asboth2016} and graphene nanoribbons,~\cite{delplace2011} we here proceed 
with a numerical calculation of the invariant. Computationally speaking, the Zak phase involves the calculation of wavefunction amplitudes with some 
arbitrary gauge introduced by the diagonalization algorithm. In consequence, their possible outcomes, i.e. $\mathcal{Z}_\alpha = 0$ or $\pi$, do not 
fully determine the band topology. However, one can infer its topology from the variation of $\mathcal{Z}_\alpha$ with respect to a reference case in 
which such a phase is known. In the previous sections, we concluded that for very high phonon energies, given by the condition 
$\hbar\omega>6\gamma_0$, there are no band crossings, and in consequence the zero-phonon replica remains unperturbed. This high-frequency limit 
represents our reference case, where the Zak phase is well known for all band replicas.

\begin{table}
\def\arraystretch{1.5}
\setlength{\tabcolsep}{0.75em}
\begin{tabular}{|c||c|c|c|c||c|c|c|c|} 
\hline
\multirow{2}{*}{$q$} & \multicolumn{4}{c||}{$\hbar\omega=5\gamma_0$} & \multicolumn{4}{c|}{$\hbar\omega=3\gamma_0$} \\ \cline{2-9}
   & 1     & 2     & 3     & 4     & 1     & 2     & 3     & 4     \\ \hline \hline
1  & 0     & $\pi$ & $\pi$ & 0     & 0     & $\pi$ & $\pi$ & 0     \\ \hline
2  & 0     & $\pi$ & $\pi$ & 0     & 0     & $\pi$ & $\pi$ & 0     \\ \hline
3  & 0     & $\pi$ & $\pi$ & 0     & 0     & $\pi$ & $\pi$ & 0     \\ \hline
4  & 0     & $\pi$ & $\pi$ & 0     & 0     & $\pi$ & $\pi$ & 0     \\ \hline
5  & 0     & $\pi$ & $\pi$ & 0     & 0     & $\pi$ & $\pi$ & 0     \\ \hline
6  & 0     & 0     & 0     & 0     & 0     & $\pi$ & $\pi$ & 0     \\ \hline
7  & 0     & 0     & 0     & 0     & 0     & $\pi$ & $\pi$ & 0     \\ \hline
8  & $\pi$ & $\pi$ & $\pi$ & $\pi$ & $\pi$ & 0     & 0     & $\pi$ \\ \hline
9  & $\pi$ & $\pi$ & $\pi$ & $\pi$ & $\pi$ & 0     & 0     & $\pi$ \\ \hline
10 & $\pi$ & $\pi$ & $\pi$ & $\pi$ & $\pi$ & $\pi$ & $\pi$ & $\pi$ \\ \hline
11 & $\pi$ & $\pi$ & $\pi$ & $\pi$ & $\pi$ & $\pi$ & $\pi$ & $\pi$ \\ \hline
\end{tabular}
\caption{Zak phases for the four bands belonging to the $N_y=11$ eigenmodes in an ac-GNR for the cases $\hbar\omega = 5\gamma_0$ and 
$\hbar\omega = 3\gamma_0$. In both cases we used $\gamma_x = 0.1 \gamma_0$.}
\label{tab:1}
\end{table}

\begin{figure*}[!ht]
\includegraphics[width=\textwidth]{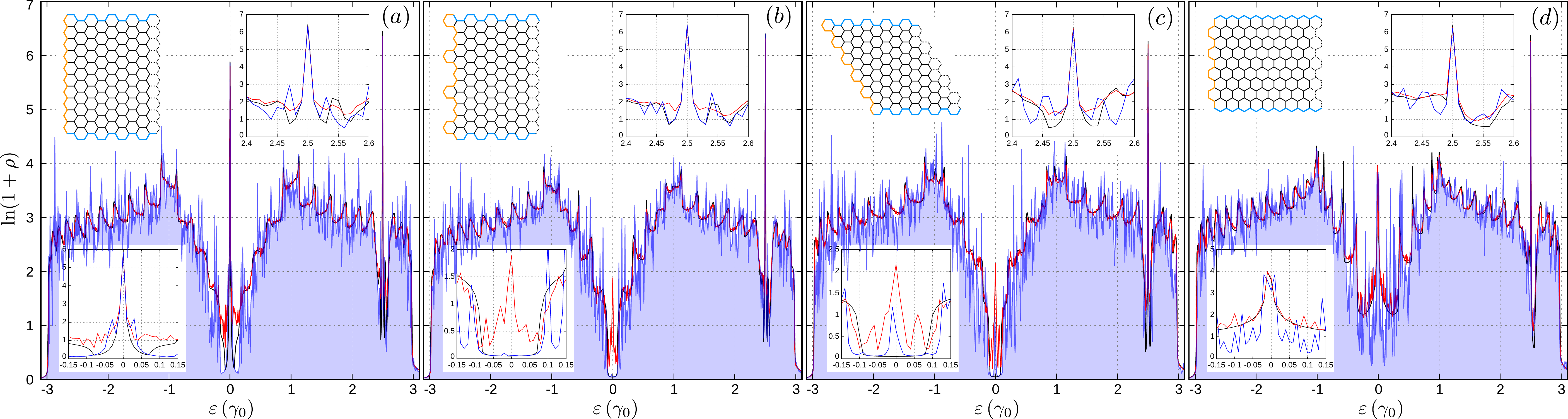}
\caption{Zero-phonon LDoS evaluated around the edge of different ribbon geometries including vacancy disorder of $0.5\%$. The evaluation region 
consists of 10 transversal layers of carbon atoms starting from the left edge of the ribbon. Black lines show the case without vacancies for 
comparison reasons, blue lines (shaded area) show a single disorder realization, and red lines show an ensamble average over $200$ disorder 
realizations. The used geometries are (see schemes): (a) armchair-zigzag ($N_y=11 \rightarrow L_y=25.82\ams$), (b) armchair-Klein 
($N_y=11 \rightarrow L_y=25.82\ams$), (c) armchair-armchair ($N_y=12 \rightarrow L_y=20.91\ams$), and (d) zigzag-armchair 
($N_y=12 \rightarrow L_y=22.01\ams$). The insets in each panel are zoom areas around $\varepsilon = 0$ (bottom) and 
$\varepsilon = \hbar\omega/2$ (top). We used two phonon replicas and the other parameters are: 
$\hbar\omega = 5\gamma_0$, $\gamma_x = 0.1 \gamma_0$.}
\label{fig:5}
\end{figure*}

We show in Table~\ref{tab:1} the Zak phases corresponding to the four bands belonging to replicas $n=0$ and $n=1$ for the phonon energies 
$\hbar\omega = 5 \gamma_0$ and $\hbar\omega = 3\gamma_0$. As can be seen from the table, in this example the role of the interaction is to open a gap 
between bands 2 and 3, and adding a factor $\pi$ to the band's cumulative phase. For $\hbar\omega=5\gamma_0$ this implies that modes with 
$q=1,\dots,5$ host vibration induced topological states between bands 2 and 3, while modes with $q=8,\dots,11$ host native topological states in 
the gap between bands 1 and 2 (related with the zero-phonon replica) and between bands 3 and 4 (related with the $n=1$ phonon replica). For 
$\hbar\omega=3\gamma_0$, the band crossing condition is fulfilled for $q=1,\dots,9$. For these modes, the Zak phase for the lowest energy band 
(valence band of the $n=0$ replica) is equal to $\pi$, which means that native edge states are present. However, more interesting is the case of the 
$q=8,9$ modes, where the topological invariant for the second lowest energy band is equal to zero (cumulative Zak phase of $\pi$), which implies that 
these modes host both native and vibrational induced topological states.

We therefore notice that the condition for the formation of interaction induced topological states is the band crossing, while the 
native states appear when the intracell hopping $\gamma_{q,\mathrm{a}}^{(0)}$ is smaller than the intercell hopping $\gamma_0$.~\cite{asboth2016} As 
we shall see next, this difference in the formation of the native and interaction-induced topological states brings with it important consequences 
when evaluating the robustness of such states against changes in the ribbon geometry and the introduction of several types of disorder.

\subsection{Ribbon geometries and robustness against disorder}

So far we have been discussing the effects of the e-ph interaction on a particular ribbon geometry where the eigenmodes remain decoupled even 
in the presence of the vibration. It therefore becomes natural to ask whether the topological states survive in other geometries and, in turn, if they 
are robust against different types of disorder which might couple these eigenmodes. In this section we provide an answer to these questions by 
analysing the LDoS along the ribbon edges for different geometries and by incorporating either vacancy or impurity disorder.

In Fig.~\ref{fig:5} we show the LDoS evaluated around the edge region for different GNRs geometries. As in Fig.~\ref{fig:3}, we plot the quantity 
$\ln(1+\rho)$ as to compensate the peaks height from the rest of the data. In these examples we considered randomly-generated vacancy disorder of 
$0.5\%$ over the complete sample. In all panels, we show the zero-phonon LDoS without disorder (solid black), a single disorder realization (shaded 
blue area), and an ensamble average over $N = 200$ realizations (red line). The insets in each panel show zoom regions around $\varepsilon = 0$ 
(bottom inset) and $\varepsilon = \hbar\omega/2$ (top inset) to help visualization, while the schemes describe the type of ribbon geometry, 
characterized by longitudinal (cyan) and transversal (orange) borders. In all cases we used $N_x \sim 1000$ to ensure a ribbon length much larger than 
the localization lengths of all edge states.

Comparing the panels in Fig.~\ref{fig:5} we first notice that, in absence of vacancy disorder, the LDoS peak in $\varepsilon = 0$ can be present or 
not depending on the ribbon geometry. This is illustrated by the black lines in the figure (see bottom inset in each panel), where panels (b) and (c) 
show no peak at this energy, while in panels (a) and (d) such a peak is certainly visible. The peak in $\varepsilon = \hbar\omega/2$, on the other 
hand, is present in all panels. This important result allows for a clear distinction between the $\varepsilon = 0$ and the 
$\varepsilon = \hbar\omega/2$ edge states. While the native states may appear or not depending on the particular ribbon geometry, the presence of 
topological states induced by the e-ph interaction is ensured by the crossing of conduction and valence bands belonging to different phonon replicas. 
For this reason we believe the native states are rather marginal: Although they admit a topological characterization through the Zak phase, their 
emergence is strictly determined by the ribbon's geometry.

\begin{figure*}[!ht]
\includegraphics[width=0.95\textwidth]{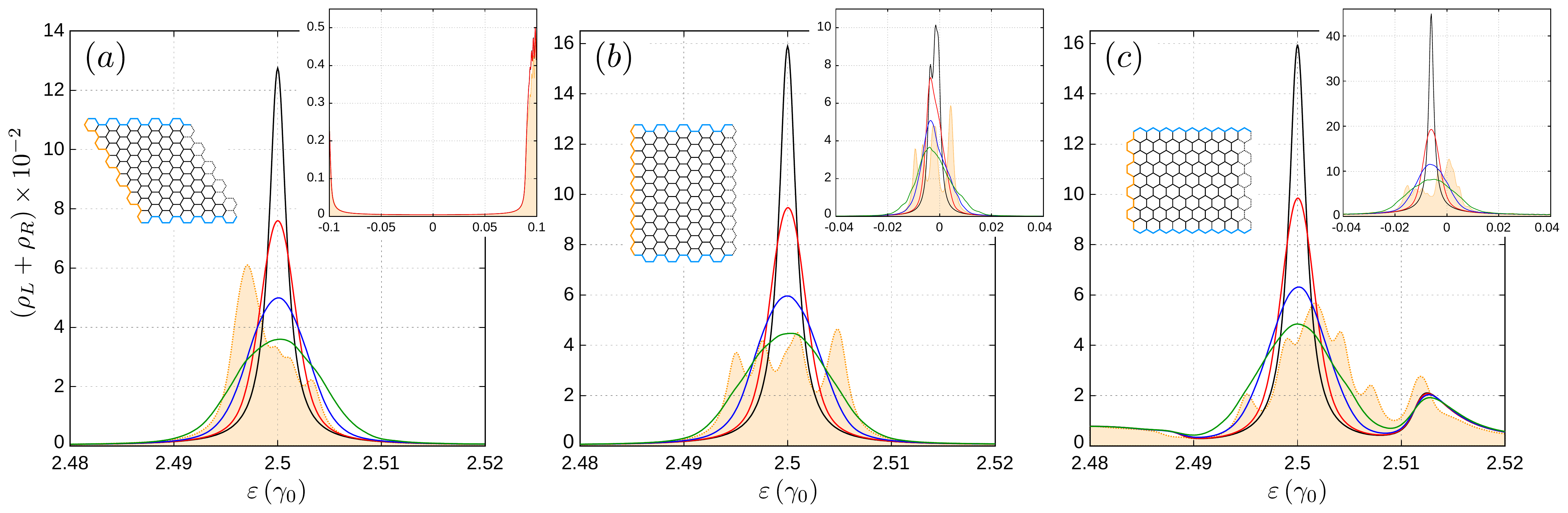}
\caption{Impurity disorder effects in the zero-phonon LDoS for three ribbon geometries: (a) armchair-armchair ($N_y = 12 \rightarrow L_y=20.91\ams$), 
(b) armchair-zigzag ($N_y = 11 \rightarrow L_y=25.82\ams$), and (c) zigzag-armchair ($N_y = 12 \rightarrow L_y=22.01\ams$). In all panels, we 
evaluated the LDoS along 40 transversal layers of carbon atoms from the left border and other 40 lines from the right border. Solid lines show the 
average LDoS over 250 disorder realizations, where we used $W = 0.01 \gamma_0$ (red), $0.02 \gamma_0$ (blue), and $0.03 \gamma_0$ (green). Black 
lines correspond to the case without disorder, while orange dotted lines (shaded) show a single realization for $W = 0.03 \gamma_0$. Main panels 
sweep the e-ph induced bandgap region, centered at $\varepsilon = \hbar\omega/2$, while the insets show the Dirac point region, centered at 
$\varepsilon = 0$. The other parameters coincide with those of Fig.~\ref{fig:5}.}
\label{fig:6}
\end{figure*} 

When we include a $0.5\%$ concentration of vacancies along the full sample, the main structure of the LDoS holds (blue lines), though 
it obviously displays a noisy pattern around the LDoS without disorder (black lines). Such a perturbative behavior can be tested by taking an 
ensamble average over several ribbon samples with the same vacancy concentration. The red lines show the LDoS averaged over $N = 200$ realizations, 
and superimpose the LDoS without disorder along almost the entire spectrum. However, looking closer at $\varepsilon = 0$ in Figs.~\ref{fig:5}(b) and 
(c) (see bottom insets) we notice that, in average, the zero energy peak returns when vacancy disorder is included. The fact that the average LDoS 
(red) in Fig.~\ref{fig:5}(b) shows a peak at $\varepsilon = 0$ while the trial LDoS for a single realization (blue) shows no peak, and that this peak is
present in both cases in Fig.~\ref{fig:5}(c) indicates that this is an intermittent effect: In those cases where the LDoS shows no central peak 
without disorder, when including disorder this peak may appear for some vacancy configurations. In fact, these peaks have nothing to do with the 
native edge states appearing in Figs.~\ref{fig:5}(a) and (d), but these are related to localized states that surround the vacancies in the 
sample.~\cite{pereira2006} Thus depending on the presence (or absence) of vacancies nearby the region where the LDoS is being evaluated one can see 
(or not) a peak in the LDoS. In the considered examples of Fig.~\ref{fig:5}(b) and (c), the region where the LDoS was evaluated involves $\sim$100 
carbon atoms, and since the vacancy concentration is $0.5\%$, one expects $\sim$0.5 vacancies in this region, so the chances of observing a peak in 
this region are one in two. Obviously, since the chances to have one or more localized states due to the presence of vacancies within the evaluation 
region grow with the vacancy concentration, we expect a simple relation between this quantity and the average height of the central peak. Having 
understood the role of disorder in the central peak of Figs.~\ref{fig:5}(b) and (c), we now observe that the shape and intensity of the central peak 
in Figs.~\ref{fig:5}(a) and (d) change little when including disorder, meaning that the native states, if present, are robust against moderate vacancy 
disorder. For the e-ph induced topological states we can arrive to the same conclusion, as the peaks in $\hbar\omega/2$ are all the same, regardless 
of the vacancy disorder.

We now investigate the role of impurity disorder on the ribbon's LDoS. This is modeled through a random variation of the on-site energies within the 
range $[-W,W]$. This means that the pure electronic Hamiltonian in Eq.~(\ref{eq:Ham_op}) is replaced by
\begin{equation}
\hat{H}_\mathrm{el} = \sum_i \epsilon_i \hat{c}_i^\dag \hat{c}_i - \sum_{\braket{i,j}} \gamma_0 \hat{c}_i^\dag \hat{c}_j,
\label{eq:Ham_dis}
\end{equation}
with $-W \leq \epsilon_i \leq W$ the random on-site energy. In Fig.~\ref{fig:6} we show the LDoS around the e-ph band crossing point, centered at 
$\varepsilon = \hbar\omega/2$, together with the LDoS around the Dirac point at $\varepsilon = 0$ (insets). In these examples we evaluate the 
zero-phonon replica LDoS over 40 transversal layers of carbon atoms for both left and right borders, considered as mirror images each other. This was 
done for three ribbon geometries (see schemes in each panel), and we used $W = 0.01 \gamma_0$ (red), $0.02 \gamma_0$ (blue), and $0.03 \gamma_0$ 
(green). In these examples we calculated the average LDoS over 250 disorder realizations. Black lines exhibit the case without disorder 
($W = 0$), while dotted orange lines (shaded area) illustrate the case of a single disorder realization for $W = 0.03 \gamma_0$.

For the shown ribbon geometries, we can see that the e-ph induced LDoS peak at $\hbar\omega/2$ (dotted orange) now splits out into several peaks 
lying within the bandgap region. To understand why this type of disorder produces such an effect, first notice that the LDoS peak at $\hbar\omega/2$ 
without disorder (black lines) can be decomposed into several peaks, each one belonging to a topological edge state. The position of these peaks 
depends on the average of the on-site energies around the region where the probability density is finite. If we imagine the impurity disorder [first 
term in Eq.~(\ref{eq:Ham_dis})] as a perturbation, and $\psi_k(\bm{r}_i)$ represents the wavefunction amplitude of the (unperturbed) e-ph induced 
topological state $k$ at position $\bm{r}_i$, then the energy $\varepsilon_k$ (and thus the peak position in the LDoS) will depend on the on-site 
energies as: 
\begin{equation}
\varepsilon_k \simeq \frac{\hbar\omega}{2}+\sum_i \epsilon_i |\psi_k(\bm{r}_i)|^2,
\end{equation}
which can be interpreted as the original position (i.e. without disorder), plus the $k$-state weighted average of the on-site energies. As for the 
considered ribbon sizes the topological edge states have finite weight over a small number of sites, the last term in the above equation may not 
vanish in general. In fact, this quantity tends to increase with the degree of disorder $W$. This is reflected as a broadening of the averaged LDoS 
peaks (solid red, blue, and green) when $W$ increases. Importantly, the area below the LDoS peak remains always constant, meaning that the number of 
topological states in the sample is independent of disorder. Of course, though $W$ does not change the number of topological states, for larger 
$W$ values these states may be located in energy regions outside the overall gap, thus difficulting a clear separation between localized and extended 
states.

The same disorder-induced peak broadending can be observed in the central region around $\varepsilon = 0$ (see insets), but it is important 
to notice that, as in the vacancy disorder case, the native states can be present or not depending on the ribbon geometry. Additionally, due to the 
zoom factor, one can observe that the average peaks are not perfectly centered at $\varepsilon = 0$, but slightly shifted to the left. This is not 
related to disorder (the black line peaks are also shifted) but a second order effect in the coupling between the zero- and one-phonon replica bands.

\section{Summary and final remarks.}

In summary, we have shown that novel and robust states of topological origin form as a consequence of the electron-phonon interaction in graphene 
nanoribbons. This study, based on a specific model for the electron-phonon interaction given by a stretching mode in graphene nanoribbons, serves as a 
proof of concept. The topological states form at the center of a bandgap (induced by the same interaction) located at half the phonon energy above the 
charge neutrality point. This was confirmed in several ribbon geometries and for vacancy and impurity disorder configurations. While both the native 
and the e-ph induced states were characterized through the Zak phase, and shown to be robust against disorder, the native states only appear in some 
specific ribbon geometries. Conversely, for a non-negligible e-ph interaction, the presence of the vibration induced topological states is guaranteed 
as long as the phonon energy does not exceed the typical band width, i.e. $\hbar\omega<6\gamma_0$. Such a condition provides the required band 
inversion between the first two phonon replicas.

Interestingly, this physics happens in our case even when the phonons do not break time-reversal symmetry, similar to a driven one-dimensional 
topological insulator.~\cite{dallago2015} In two dimensional systems, however, the e-ph induced bandgap finally closes for some particular mode 
$\bm{k}$ (see discussion on the $N_y \rightarrow \infty$ limit around Fig.~\ref{fig:2}) and one should break TRS to restore the gap (as it is required 
for light to induce Floquet topological states in the same material). In this sense, the recent observation of chiral phonons in two-dimensional 
materials~\cite{zhu2018} may open a promising way for studying electron-phonon induced topological phase transitions.

\vspace{0.5cm}

\noindent
\textit{Acknowlegdments.--} This work was supported by Consejo Nacional de Investigaciones Cient\'ificas y T\'ecnicas (CONICET), Secretar\'ia de 
Ciencia y Tecnolog\'ia -- Universidad Nacional de C\'ordoba (SECYT--UNC). HLC and JSL are members of CONICET. LEFFT acknowledges funding from Program 
“Instalación Académica” 2016 of the Faculty of Physical and Mathematical Sciences of the University of Chile and FondeCyT (Chile) under project number 
1170917.

\appendix
\section{Vibration induced bandgaps}
\label{app:1}

Let us think of two replicas (zero- and one-phonon) of a dimer chain for a particular eigenmode $q$ (see Fig.~\ref{fig:1}). Each dimer chain 
develops a valence and a conduction band, and the energy difference between the two replicas is $\hbar\omega$. Let us suppose that $\hbar\omega$ is 
small enough such that the $n=0$ conduction band and the $n=1$ valence band cross. If $k$ denotes the Bloch quasimomentum, there are two $k$-values 
($\pm k^\ast$) where the band crossing occurs. If we focus on one of these points, say $k^\ast$, we have two $k$-states at the same energy, 
each one belonging to one of the two replicas. This degeneration gets removed by the electron-phonon interaction, which in our case corresponds to 
the coupling between the replicas, and yields the bandgap opening. In this appendix we estimate the gap size produced by the e-ph interaction.

When using the eigenmode decomposition, the hopping term between sites depends on the mode $q$ we are looking at, with a factor 
$\cos[q\pi/(2N_y+1)]$, and $q = 1, \ldots, N_y$. We take the cosine argument as a continuous variable $x$, within the range $0<x<\pi/2$, and we 
simplify this analysis by truncating the full Fock space so we only keep the $n=0$ and $n=1$ replicas. Considering the bulk situation (i.e. an 
infite long dimer chain) we can use Bloch theorem and obtain the following Hamiltonian:
\begin{equation}
\bm{H}_q = \left(\begin{array}{cccc}
0               & v_q^{(0)} & 0                & v_q^{(1)}  \\
\bar{v}_q^{(0)} & 0         & \bar{v}_q^{(1)}  & 0          \\
0               & v_q^{(1)} & \hbar\omega      & v_q^{(0)}  \\
\bar{v}_q^{(1)} & 0         & \bar{v}_q^{(0)}  & \hbar\omega
\end{array}\right), 
\end{equation}
where
\begin{eqnarray}
v_q^{(n)} = |v_q^{(n)}| \exp\left[i\varphi_q^{(n)}\right] = -\gamma_{q,\mathrm{a}}^{(n)} - \gamma_{q,\mathrm{b}}^{(n)} e^{-ika},
\label{eq:veff}
\end{eqnarray}
with $a=3a_0/2$ the unit cell length, and the bar standing for complex conjugation. Since we are only interested in the $n=0$ conduction and $n=1$ 
valence bands, we can reduce even more this Hamiltonian. To do so, we first diagonalize the $2 \times 2$ block matrices in the diagonal. As they 
commute each other, we can diagonalize the $n=0$ block, and obtain the energies $\varepsilon = \pm |v_q^{(0)}|$. Similarly, for the 
$n=1$ block we obtain $\varepsilon = \hbar\omega \pm |v_q^{(0)}|$. The next step is to write the Bloch Hamiltonian in this new basis, such that it 
allows the proper truncation
\begin{equation}
  \tilde{\bm{H}}_q = \left(\begin{array}{cc}
 | v_q^{(0)} | & -i |v_q^{(1)}| \sin(\Delta\varphi) \\
 i |v_q^{(1)}| \sin(\Delta\varphi) & - | v_q^{(0)} | + \hbar \omega  
\end{array}\right),
\label{eq:Ham_eff}
\end{equation}
where $\Delta\varphi = \varphi_q^{(1)}-\varphi_q^{(0)}$. This Hamiltonian has the following eigenvalues
\begin{equation}
\varepsilon_{q,\pm} = \frac{\hbar\omega}{2} \pm \sqrt{\left(\frac{\hbar\omega}{2}-|v_q^{(0)}|\right)^2 
+ \left(|v_q^{(1)}|\sin(\Delta\varphi)\right)^2}.
\end{equation}
With these expressions, we found the new eigenenergies for $k$ close to the band crossing point. We can, in turn, particularize to the point in which
this band crossing occurs and obtain an expression for the size of the gap $\Delta(x)$. Recalling that the band crossing takes place at 
$\varepsilon=\hbar\omega/2$, we take $|v_q^{(0)}|$ equal to this energy, and using the definitions for $v_q^{(n)}$ given in Eq.~(\ref{eq:veff}), we 
obtain
\begin{equation}
\Delta(x) = 2 \sqrt{\frac{9 \gamma^2_x \cos^2 x \sin^2 ka}{1 + 4 \cos^2 x + 4 \cos x \cos ka}}. 
\end{equation}
By finding the $k$-value for which $|v_q^{(0)}|=\hbar\omega/2$ is satisfied, and defining the adimensional parameter $\eta=\hbar\omega/4\gamma_0$, we
obtain that the size of the e-ph interaction induced bandgap is given by
\begin{equation}
\Delta(x)  = \frac{3 \gamma_x}{\eta} \sqrt{\left(\eta_{+}^2-\cos^2 x\right)\left(\cos^2 x-\eta_{-}^2\right)},
\end{equation}
where $\eta_\pm = \eta \pm 1/2$. The size of the gap will depend, therefore, on this parameter $\eta$ and the particular eigenmode $q$ 
(through the $x$-variable) we are looking at. The square root argument defines the band crossing regimes, as this quantity needs to be always
positive. This yields the condition
\begin{equation}
  |\eta_{-}| < \cos x < \eta_{+},
\end{equation}
which is equivalent to Eq.~(\ref{eq:ranw}) of the main text.

\bibliographystyle{apsrev4-1_title}
\bibliography{cite}

\end{document}